\documentclass[aps, prl,nofootinbib,tightenlines, twocolumn,floatfix]{revtex4}

 \usepackage[dvips,final]{graphicx}
  \usepackage{amssymb}
   \usepackage{amsmath}
    \usepackage{amsfonts}
     \usepackage{epsfig}
      \usepackage{bm} % bold math
\makeatletter\AtBeginDocument{\let\@elt\relax}\makeatother

\usepackage{mathrsfs}
\usepackage{slashed}

\usepackage[colorlinks = true,
linkcolor = blue,
urlcolor  = blue,
citecolor = blue,
anchorcolor = blue]{hyperref}

\begin{document}

\title{Entanglement Viscosity: from Unitarity to Irreversibility in Accelerated Frames}

\author{G. Yu. Prokhorov$^{\, a,b}$}
\email{prokhorov@theor.jinr.ru}

%\author{O. V. Teryaev$^{\, a,b}$}
%\email{teryaev@jinr.ru}

%\author{V. I. Zakharov$^{\, b,a}$}
%\email{vzakharov@itep.ru}

\affiliation{$^{a\,}$Joint Institute for Nuclear Research, Joliot-Curie 6, Dubna, 141980, Russia}
\affiliation{$^{b\,}$NRC Kurchatov Institute, Moscow, Russia}
%\affiliation{Pacific Quantum Center,
%Far Eastern Federal University, 10 Ajax Bay, Russky Island, Vladivostok 690950, Russia}

\begin{abstract}
We demonstrate that the unitarity of quantum field theory, through the positivity of spectral functions, underlies thermodynamic irreversibility for a subsystem separated by a horizon, in direct analogy with the irreversibility of renormalization-group flows. To this end, we explicitly find the shear and bulk viscosities -- the entanglement viscosities -- for thermal radiation in Rindler space using the universal spectral representation. A direct consequence of the obtained general formulas is the relationship between the acceleration-induced shear viscosity in flat space and the conformal quantum anomaly in curved space, pointing to a possible novel probe of the conformal anomaly in systems with extreme acceleration. Moreover, for conformal field theories, we explicitly show that globally entanglement viscosity saturates the Kovtun-Son-Starinets bound.
\end{abstract}

\maketitle

%================
\section{Introduction}
\label{sec_intro}
%================

The emergence of thermodynamic irreversibility from unitary microscopic equations is a fundamental question \cite{Prigogine1978-uh}. A significant new element in this problem was introduced by black hole physics. Indeed, an object falling into a black hole cannot (or rather, can, but with a very low probability) return from below the horizon. According to modern concepts, this irreversibility is a fundamental manifestation of thermodynamic \cite{Witten:2024upt}.

In this paper, we consider a simple example of a ``flat-space'' black hole, described by the Rindler horizon of an accelerated reference frame \cite{Unruh:1976db, Witten:2024upt, Bisognano:1976za}. By studying viscosity of the thermal radiation in the Rindler space, we directly demonstrate that the unitarity of the total theory not only is compatible with irreversibility in the subsystem, but in fact determines the direction of its thermodynamic evolution toward increasing entropy. This makes the situation analogous to the irreversibility of renormalization group flows, where unitarity also plays a key role \cite{Zamolodchikov:1986gt, Cappelli:1990yc, Komargodski:2011vj}.

The study of the deep connection between viscosity and black hole physics has a long history. A significant early achievement was the membrane paradigm, where this connection is most directly evident \cite{Damour:1979wya, Parikh:1997ma}. The development of this concept in the context of the holographic approach and string theory led to the remarkable prediction of the existence of a fundamental bound on the ratio of the shear viscosity $ \eta $ to the entropy density $ s $ \cite{Kovtun:2004de}
\begin{eqnarray}
\frac{\eta}{s}\geq \frac{1}{4\pi}\,.
\label{kss}
\end{eqnarray}

However, the viscosity of a black hole, or more precisely, the membrane located above it, is a purely classical effect of classical gravitational equations. An unexpected observation was that the thermal radiation of a black hole itself also exhibits viscosity (as demonstrated using the Rindler space), leading to the idea of ``entanglement viscosity'' \cite{Chirco:2010xx, Lapygin:2025zhn}. Indeed, given that entanglement leads to information loss, and in hydrodynamics the increase in entropy is associated with viscosity \cite{landau1987fluid}, one might expect that entanglement should lead to viscosity even for systems of free fields. The concept of entanglement viscosity also finds an intriguing parallel with the irreversibility induced by quantum measurement in emergent polarization phenomena \cite{Teryaev:2022mpe}.

The main result of the present paper is a direct calculation of the dissipative transport coefficients for Unruh radiation within the framework of the linear response theory, using the universal spectral representation \cite{Cappelli:1990yc, Smolkin:2014hba}. As a result, we explicitly express the shear $ \eta $ and bulk $ \zeta $ viscosity of the thermal radiation in the Rindler space for a wide class of field theories (including theories with massive and interacting fields in space of any dimension) in terms of the spin-0 $ c^{(0)}(\mu) $ and spin-2 $ c^{(2)}(\mu) $ spectral densities of the vacuum correlator of two energy-momentum tensors
\begin{eqnarray} \nonumber
\eta (\rho) &=& k_d \rho \int_0^{\infty}d\mu\, c^{(2)}(\mu)\mu^2 K_0(\mu\rho)\,, \\
\zeta (\rho) &=& \frac{2k_d \rho}{(d-1)^2}  \int_0^{\infty}d\mu\, c^{(0)}(\mu)\mu^2 K_0(\mu\rho)\,,
\label{main}
\end{eqnarray}
where the constant coefficient $ k_d=\frac{\pi^{d/2}}{(d+1)2^{d-1}\Gamma(d)\Gamma(d/2)} $ is determined by the spacetime dimension $ d $, and $ \rho $ is the distance from the Rindler horizon. In two dimensions, only the spectral density $ c^{(0)}(\mu) $ exists, from which the monotonous c-function can be constructed \cite{Zamolodchikov:1986gt, Cappelli:1990yc}\footnote{There are indications that the function $ c^{(0)}(\mu) $ may be essential in generalizing the c-theorem above two dimensions \cite{Solodukhin:2013yha, Cappelli:1990yc}.}.

Considering that the unitarity of quantum field theory fixes the sign of the spectral functions $ c^{(0)}(\mu),c^{(2)}(\mu) \geq 0 $, and the Bessel function is positive $ K_0(x) > 0 $ for $ x>0 $, we instantly conclude from (\ref{main}), that unitarity provides the non-negativity of the shear and bulk viscosity $ \eta,\zeta\geq 0 $. But the non-negativity of the viscosities is precisely the reason ensuring the thermodynamic irreversibility $ \partial_{\mu}s^{\mu} \geq 0 $, where $ s^{\mu} $ is the entropy current \cite{landau1987fluid}. Thus, we conclude that the unitarity of quantum field theory fixes the direction of thermodynamic evolution for a subsystem separated by a horizon. 

Since the same positivity of the $ c^{(0)}(\mu)$ function simultaneously determines (in 2 dimensions) the irreversibility of the renormalization group flows  \cite{Zamolodchikov:1986gt, Cappelli:1990yc, Komargodski:2011vj}, then in the example under consideration both irreversibilities arise simultaneously and in a similar way \cite{Grozdanov:2016fkt}.

We also show, considering a special case of conformal field theory, that the ``global'' entanglement viscosity of the Unruh radiation, integrated from a stretched horizon to infinity, saturates the bound (\ref{kss}), which generalizes the analysis of \cite{Chirco:2010xx, Lapygin:2025zhn} to an arbitrary conformal field theory. At the same time, we show that previously known results for free massless fields with spins 0, 1/2, and 1 are exactly reproduced from (\ref{main}).

We use the system of units $ e=\hbar=c=k_B=1 $, and the metric signature $ (+,-,-,-) $.

%================
\section{Entanglement viscosity from spectral functions}
\label{sec_main}
%================

Let us calculate the viscosity of thermal radiation in the d-dimensional Rindler space in the general case. We assume that the temperature is equal to the Unruh temperature, $ T_U = \alpha/2\pi $, where $ \alpha $ is the absolute value of the 4-acceleration of the reference frame, that is, the system is in the Minkowski vacuum state \cite{Becattini:2017ljh}. 

For $ T=T_U $ formulas (for example, for Green functions) in Rindler space can be derived from their Minkowski counterparts by an appropriate change of coordinates \cite{Unruh:1983ac}. So, let us first consider the d-dimensional flat Minkowski space, of which the Rindler space is a subsystem.  In \cite{Cappelli:1990yc} (see also \cite{Smolkin:2014hba}), a universal spectral representation for the correlator of two energy-momentum tensors was introduced, based on general symmetry considerations\footnote{In \cite{Cappelli:1990yc} Euclidean case was considered, and we analytically continue the spectral representation to non-Euclidean Minkowski spacetime \cite{Osterwalder:1973dx}.}
\begin{eqnarray}
&&\langle 0| \hat{T}_{\alpha\beta} (x) \hat{T}_{\rho\sigma} (x')|0\rangle_M = \nonumber\\ &&=\frac{A_d}{(d-1)^2} \int_0^{\infty} d\mu\, c^{(0)}(\mu) \Pi^{(0)}_{\alpha\beta,\rho\sigma}(\partial') G_d (x-x',\mu) \nonumber\\
&&+\frac{A_d}{(d-1)^2} \int_0^{\infty} d\mu\, c^{(2)}(\mu) \Pi^{(2)}_{\alpha\beta,\rho\sigma}(\partial') G_d (x-x',\mu)\,,\quad \label{spectral}
\end{eqnarray}
according to which all the differences between the theories come down to different spectral densities $ c^{(0)}(\mu) $ and $ c^{(2)}(\mu) $. Here $ \partial'=\partial/\partial x' $, $ A_d= \frac{\Omega_{d-1}}{(d+1)2^{d-1}} $, $ \Omega_{d-1}=\frac{2 \pi^{d/2}}{\Gamma(d/2)} $, averaging is performed over the Minkowski vacuum $\langle 0| ... |0\rangle_M $, and the notations are introduced
\begin{eqnarray}
&& \Pi^{(0)}_{\alpha\beta,\rho\sigma} (\partial') = \frac{1}{\Gamma(d)} S_{\alpha\beta} S_{\rho \sigma}\,, \nonumber\\
&& \Pi^{(2)}_{\alpha\beta,\rho\sigma} (\partial') = \frac{d-1}{2\Gamma(d-1)} (S_{\alpha\rho} S_{\beta \sigma}+S_{\alpha\sigma} S_{\beta \rho}\nonumber\\
&&-\frac{2}{d-1}S_{\alpha\beta} S_{\rho \sigma})\,, \nonumber\\
&& S_{\alpha\beta}(\partial')=\partial_{\alpha}'\partial_{\beta}'- \eta_{\alpha\beta} \partial'^2 \,.
\label{pps}
\end{eqnarray}
According to (\ref{pps}) the correlator is generally expressed through a scalar massive propagator constructed as an analytic continuation of the Euclidean propagator
\begin{eqnarray}
&& G_{d} (x-x',\mu) = \lim_{\varepsilon\to 0^+} G^{E}_{d} (x-x',\mu)|_{t\to i t+\varepsilon}\,,
\nonumber\\
&&G^{E}_{d} (x-x',\mu)=\int \frac{d^d p}{(2\pi)^d}\frac{e^{ip(x-x')}}{p^2 + \mu^2}\,,
\label{prop}
\end{eqnarray}
where the second formula uses the Euclidean signature accordingly. The prescription with $ \varepsilon>0 $ corresponds to the transition to the Wightman function (without time ordering).  

The representation (\ref{spectral}) is valid for a wide class of theories, including theories with interactions and/or massive fields and with different spins. The unitarity condition for a quantum field theory in Minkowski space leads to the non-negativity condition for these spectral densities
\begin{eqnarray}
c^{(0)}(\mu) \geq 0,\quad
c^{(2)}(\mu) \geq 0\,.
\label{unit}
\end{eqnarray}

Let us now proceed directly to the calculation of viscosity in the Rindler space. Transport coefficients can be found using the Kubo formulas \cite{Landau9, Son:2008zz, Laine:2016hma, Kharzeev:2007wb}. In particular, for the shear viscosity in space with some metric $ g_{\mu\nu} $\footnote{Note that Kubo formulas are often written in the literature without the factor $ \sqrt{g^{00}} $, since the case $ g^{00}=1 $ is often considered.}
\begin{eqnarray}
&& \eta(x') = \text{Im}\lim_{\omega \to 0^+} \frac{i}{\sqrt{g^{00} (x')} \omega} \int_{-\infty}^{\infty} dt\, \theta(t-t') \nonumber\\
&& \cdot\,  e^{i\omega (t-t')}  \int_{V}d^{d-1} x\, \sqrt{-g(x)}\,  \langle [\hat{T}_{12} (x), \hat{T}_{12} (x')]\rangle\,,
\label{kuboeta}
\end{eqnarray}
and a similar formula for bulk viscosity
\begin{eqnarray}
&& \zeta(x') = \frac{1}{(d-1)^2}\text{Im}\lim_{\omega \to 0^+} \frac{i}{\sqrt{g^{00} (x')} \omega} \int_{-\infty}^{\infty} dt\, \theta(t-t') \nonumber\\
&& \cdot\, e^{i\omega (t-t')} \int_{V}d^{d-1} x\, \sqrt{-g(x)}\, \langle [\hat{T}_{\alpha}^{\alpha} (x), \hat{T}_{\sigma}^{\sigma} (x')]\rangle\,,
\label{kubozeta}
\end{eqnarray}
where $ \hat{T}_{12} $ is the off-diagonal spatial component of the energy-momentum tensor, and $ V $ is the space volume of the system.

We are essentially interested in the viscosity of the Minkowski vacuum. And as with the Unruh effect \cite{Unruh:1976db}, a key role is played by whether we are considering an inertial or accelerated observer. In the former case, due to translational invariance, the viscosity does not depend on the coordinates and is determined by the infrared behavior of the spectral functions, in particular, it turns out to be equal to zero for massive free fields (a more detailed discussion of this issue is beyond the scope of this letter).

However, the result changes dramatically if we move from Minkowski coordinates $ (t,x^1,x^2,...,x^{d-2},x^{d-1}) $ to an accelerated observer and Rindler coordinates $ (\tau,x^1,x^2,...,x^{d-2},\rho) $ with a metric of the form
\begin{eqnarray}
ds^2 = \rho^2 d\tau^2-(dx^1)^2-...-(dx^{d-2})^2-d\rho^2\,,
\label{rindler}
\end{eqnarray}
where we assume that the acceleration occurs along the $ x^{d-1} $ axis and the Rindler coordinates are related to the Minkowski coordinates by the familiar formulas $t=\rho \sinh \tau$,  $ x^{d-1}=\rho \cosh \tau$. The Rindler coordinates (\ref{rindler}) cover only a part of the Minkowski space, the right Rindler wedge, and the hypersurface $ \rho=0 $  corresponds to the event horizon. Unlike the case with Minkowski space, translational invariance is violated by the introduction of the horizon - the fields live in the region $ \rho>0 $.

Let us start with the shear viscosity and consider the coordinates $ x^1 $ and $ x^2 $ as transverse coordinates in Kubo formula (\ref{kuboeta}) (the result does not depend on which pair of coordinates is chosen from $ (x^1,..,x^{d-2}) $).
We will use the fact that the Green functions in the Rindler space for $ T=T_U $ are expressed in terms of the usual vacuum Green functions in Minkowski space up to a change of coordinates \cite{Unruh:1983ac}. Thus, to calculate the viscosity in the Rindler space, we can use the representation (\ref{pps}), simply by switching to Rindler coordinates.

Taking into account the translational invariance with respect to the coordinates $ x^1,..., x^{d-2} $, it can be shown that $ \eta $ is expressed only through $ c^{(2)}(\mu) $, and in $ \Pi^{(2)}$ only the contribution with the square of the d'Alembertian remains. Using the equation for the Wightman function \cite{Bogoliubov1959-sq} $ \square G_d(x-x') = -\mu^2 G_d(x-x') $ and the descent method \cite{Solodukhin:2011gn}, we can also immediately integrate over the transverse coordinates $ \int_{-\infty}^{\infty}d^{d-2}x $. 
%The remaining two-dimensional integral over momentum leads to the Bessel function $ K_0 $. 
As a result, we obtain
\begin{eqnarray}
&& \eta(\rho') = k_d \rho'\,\int d\mu\, \mu^4 c^{(2)}(\mu)\, \text{Im}\,\lim_{\omega \to 0^+} \frac{i}{\omega} \int_{0}^{\infty} d\tau\ e^{i\omega \tau} \nonumber\\
&&\cdot\int_0^{\infty}d\rho \, \frac{\rho}{2\pi} \Big\{ K_0\left(\mu\sqrt{\rho^2+\rho'^2-2 \rho\rho' \cosh(\tau - i\varepsilon)}\right)\nonumber\\
&&-(\tau\to-\tau)\Big\}\,,
\label{kuboK}
\end{eqnarray}
where in the formula under the root is the standard expression for the square of the interval in Rindler coordinates. It is now convenient to move on to the momentum representation, according to the formula \cite{Prudnikov1998-xg, Moschella:2008ik}
\begin{eqnarray}\label{momentum}
&&\frac{1}{2\pi} K_0\left(\mu\sqrt{\rho^2+\rho'^2-2 \rho\rho' \cosh(\tau - i\varepsilon)}\right) = \\ 
&&=\frac{1}{\pi^2}\int_0^{\infty} d\lambda\, K_{i\lambda} (\mu \rho)K_{i\lambda} (\mu \rho') \cosh(\pi \lambda - i \lambda (\tau - i\varepsilon))\,.\nonumber
\end{eqnarray}
Then the dependence on $ \rho $, $ \rho' $, and $ \tau $ is factorized. The integral over $ \rho $ is taken analytically according to the formula \cite{Prudnikov1998-xg}
\begin{eqnarray}
\int_0^{\infty} d\rho \, \rho K_{i\lambda}(\mu \rho)
 = \frac{\pi \lambda}{2 \mu^2 \sinh(\pi \lambda/2)}\,,
\label{mom}
\end{eqnarray}
and the integral over $ \tau $ contains frequency parts of delta function \cite{Bogoliubov1959-sq}
\begin{eqnarray}
\delta_{\pm}(\alpha)=\frac{1}{2\pi}\int_0^{\infty} d\tau e^{\pm i \alpha \tau} = \frac{1}{2} \delta(\alpha)\pm \frac{i}{2 \pi} \mathrm{P}\frac{1}{\alpha} \,.
\label{delta}
\end{eqnarray}
The principal value parts $ \mathrm{P} $ drop out of (\ref{kuboK}), since we are interested in the imaginary part of the correlator. As a result, we obtain
\begin{eqnarray}
&&\eta = \frac{k_d \rho'}{4} \lim_{\omega\to 0^+} \frac{1}{\omega} \int_0^{\infty} d\mu\, c^{(2)}(\mu) \mu^2 \int_0^{\infty} d\lambda \, \frac{\lambda  K_{i\lambda}(\mu \rho')}{\sinh(\pi\lambda/2)}\nonumber\\
&&\cdot\,\Big\{e^{\pi\lambda} \delta(\omega-\lambda)+e^{-\pi\lambda} \delta(\omega+\lambda)-e^{\pi\lambda} \delta(\omega+\lambda)\nonumber\\
&&-e^{-\pi\lambda} \delta(\omega-\lambda)\Big\}\,.
\label{kubodelta}
\end{eqnarray}
Considering that the limit is taken from above $ \omega\to 0^+ $, the second and third terms in the braces in (\ref {kubodelta}) give zero. The integral over $ \lambda $ is removed by delta functions, after which the limit $ \omega \to 0^+ $ is taken. As a result, we obtain the first of the formulas (\ref{main})
\begin{eqnarray}
\eta (\rho) = k_d \rho \int_0^{\infty}d\mu\, c^{(2)}(\mu)\mu^2 K_0(\mu\rho)\,.
\end{eqnarray}
The calculation of bulk viscosity $ \zeta $ is similar. $ c^{(2)}(\mu) $ does not contribute in this case, since $ \Pi^{(2)} {{}_{\alpha}}^{\alpha}{{}_{,\beta}}^{\beta}=0$.

%================
\section{Discussion and Applications}
\label{sec_disc}
%================

%================
\subsection{From quantum-field unitarity to thermal irreversibility}
\label{secsub_irr}
%================

Thus, the entanglement viscosity is precisely determined by the spectral functions $ c^{(0)}(\mu) $ and $ c^{(2)}(\mu) $, according to formulas (\ref{main}). Moreover, the number of independent viscosity coefficients in hydrodynamics (shear and bulk), which is determined by the symmetries of the fluid (macroscopic), exactly corresponds to the number of independent spectral densities, determined by the symmetries of the Minkowski vacuum (microscopic). Thus, formulas (\ref{main}) illustrate the relationship between these two types of symmetries.

The key point is that the modified Bessel functions are positive $ K_0(x)>0 $ for $ x>0 $. And the functions $ c^{(0)}(\mu) $ and $ c^{(2)}(\mu) $ themselves are positive (\ref{unit}), as a consequence of the unitarity of the total theory in Minkowski spacetime. On the other hand, the positivity of the viscosities is a well-known consequence of the second law of thermodynamics \cite{landau1987fluid}. Thus, thermodynamic irreversibility in the case under consideration for a subsystem separated by a horizon is a direct consequence of the unitarity of quantum field theory in Minkowski space.

%================
\subsection{Viscosity and entropy from the conformal anomaly}
\label{secsub_entropy}
%================

In this subsection we demonstrate that the derived shear viscosity is a novel type of anomalous transport phenomenon, governed by the conformal anomaly. We will restrict ourselves to the case of four dimensions. The corresponding quantum anomaly has the form \cite{Duff:1993wm, Brown:1986jy}\footnote{We denote the anomaly coefficients according to \cite{Brown:1986jy}, but in the signature $ (+,-,-,-) $.}
\begin{eqnarray}
\langle \hat{T}^{\mu}_{\mu} \rangle =
a (-H + \frac{2}{3}\square R) - b  E_4 + c\, \square R\,,
\label{anom}
\end{eqnarray}
where $ a, b, c $ are numerical coefficients, $ H $ is the square of the Weyl tensor, and $ E_4 $ is the Euler density.
In the case of a conformal field theory, the spectral functions have the form \cite{Cappelli:1990yc}
\begin{eqnarray}
c^{(0)}(\mu)\sim \mu^{2} \delta(\mu)\,,\quad c^{(2)}(\mu)=  \frac{3 C_T}{4} \mu \,,
\label{conf}
\end{eqnarray}
where $ C_T $ is the conformal central charge encoding the number of degrees of freedom. For example, a theory with $ n_s $ free scalars, $ n_f $ Dirac fermions, and $ n_v $ vector fields has  $ C_T = \frac{1}{3\pi^4} (n_s +6 n_f + 12 n_v)$. On the other hand, conformal symmetry fixes the relationship between $ C_T $ and the factor $ a $ in the conformal anomaly (\ref{anom}) \cite{Osborn:1993cr, Cappelli:1990yc}
\begin{eqnarray}
C_T=\frac{640}{\pi^2} a\,.
\label{cta}
\end{eqnarray}
Substituting (\ref{conf}) into (\ref{main}), using (\ref{cta}) and also recalling that the distance to the horizon is related to the acceleration magnitude $ \rho = 1/\alpha $, we obtain the relationship between the conformal anomaly and the entanglement viscosity in the conformal field theory case
\begin{eqnarray}
\eta =  8 a  \alpha^3 \,,\quad \zeta = 0\,.
\label{etaanom}
\end{eqnarray}
A similar relationship between viscosity and the central charge was discussed within the holographic approach \cite{Kovtun:2008kx, Buchel:2008vz, Sinha:2009ev}.

Formula (\ref{etaanom}), which is a direct consequence of (\ref{main}), is confirmed by direct calculation of the entanglement viscosity for free massless fields with spins 0, 1/2, and 1 \cite{Chirco:2010xx, Lapygin:2025zhn}. Thus, the conformal quantum anomaly, an effect in curved space, determines the transport effect in flat space in an accelerated medium. This places this effect on a par with other anomalous transport effects, such as chiral magnetic (CME) and chiral vortical (CVE) effects \cite{Son:2009tf, Fukushima:2008xe, Landsteiner:2011cp} and kinematic vortical effect (KVE) \cite{Prokhorov:2022udo, Prokhorov:2022snx}, and raises the interesting question of whether the conformal anomaly, via the entanglement viscosity, could provide a non-negligible contribution to dissipative phenomena in heavy-ion collisions, particularly at the early stages, when acceleration is expected to be maximal \cite{Prokhorov:2025vak}. 

%
%This anomalous effect could be relevant for systems with extreme acceleration. For instance, using a maximal acceleration estimate of $ a \sim 1 \,Gev $ from modeling of heavy-ion collisions \cite{Prokhorov:2025vak}, we obtain a very rough estimate for shear viscosity from (\ref{etaanom}) that is comparable to the existing estimates of viscosity for the quark–gluon plasma \cite{Harris:2023tti, Singh:2025dza, Altenkort:2022yhb}. This raises the interesting question of whether entanglement viscosity could provide a non-negligible contribution to dissipative phenomena in heavy-ion collisions, particularly at the early stages of collision, when acceleration is maximal \cite{Prokhorov:2025vak}.

%In particular, there is kinematic vortical effect (KVE), which is similarly determined by the chiral anomaly in the gravitational field, but also exists in the flat space limit \cite{Prokhorov:2022udo, Prokhorov:2022snx}. Note, however, that in contrast to the case of chiral anomalies, which are thought to lead mostly to non-dissipative transport effects, the conformal anomaly, is associated with a dissipative phenomenon.

Interestingly, it is also possible to demonstrate a relationship between entropy density and conformal anomaly. There are many approaches to calculating the entanglement entropy \cite{Solodukhin:2011gn, Diakonov:2025wtt}. In particular, within relativistic spin hydrodynamics it is associated with the derivative of pressure at constant acceleration \cite{Becattini:2023ouz}
\begin{eqnarray}
s=\frac{\partial p(T,\alpha)}{\partial T}\Big|_{\alpha} \,.
\label{sbec}
\end{eqnarray}
On the other hand, the energy-momentum tensor of an accelerated medium at finite temperature, within conformal field theory, is related to the conformal anomaly (\ref {anom}) via Eq.~(4.2) from \cite {Brown:1986jy}. Using the corresponding linear relations and (\ref {sbec}), we obtain that only the anomaly coefficient $ a $ contributes to the entropy density 
\begin{eqnarray}
s(T=T_U,\alpha)=\frac{32 \pi a}{3} \alpha^3\,.
\label{sres}
\end{eqnarray}

%================
\subsection{Viscosity to entropy density ratio}
\label{secsubkss}
%================

A natural question that arises when considering shear viscosity is the relationship with the KSS bound (\ref{kss}).
To establish the relation to the KSS bound, we introduce a stretched horizon at a distance  $ l_c $ from the true horizon. This regularizes the integrals over $ \rho $ from below.

We will consider a special case of conformal field theory in four dimensions. One could start from the calculation given in the first part of the current paper, introducing regularization into the integral (\ref{mom}). However, it is useful to follow a slightly different approach and take the integral over $ \mu $ in (\ref{spectral}) from the very beginning. After that we come to the well-known expression \footnote{We also analytically continue it to the non-Euclidean space.}
\begin{eqnarray}
\langle 0| \hat{T}_{\mu\nu} (x) \hat{T}_{\alpha\beta} (x')|0\rangle_M = \frac{C_T \mathcal{I}_{\mu\nu,\alpha\beta}}{\left[(x-x')^2-i\varepsilon (x^0-{x^0}')\right]^4}\,,\quad
\label{erdm}
\end{eqnarray}
where $ \mathcal{I}_{\mu\nu,\alpha\beta} $ is the inversion tensor \cite{Erdmenger:1996yc, Smolkin:2014hba}. The subsequent steps are similar to the procedure, described in \cite{Chirco:2010xx, Lapygin:2025zhn} up to the common coefficient in the correlator $ \langle \hat{T}(x)\hat{T}(x') $. 
%We need to substitute (\ref{erdm}) into (\ref{kuboeta}), and the most nontrivial moment is the integration over the Rindler time, which reduces to finding the residue at the third-order pole. 
As a result, we obtain
%for the local shear viscosity at some distance $ \rho $ from the horizon
\begin{eqnarray}
\eta(\rho)=\frac{C_T  \pi^2 \rho \left[\rho^4+4\rho^2l_c^2-5l_c^4-4l_c^2(2\rho^2+l_c^2)\ln\frac{\rho}{l_c}\right]}{80 (\rho^2-l_c^2)^4}\,.\quad
\label{etaloc}
\end{eqnarray}
In the limit $ l_c\to 0 $ (\ref{etaloc}) exactly corresponds to (\ref{etaanom}), as it should be. Also, substituting the values of $ C_T $ for scalar, vector and Dirac massless fields,  (\ref{etaloc}) exactly corresponds to the results of \cite{Chirco:2010xx, Lapygin:2025zhn}. Thus, the local viscosity for conformal theories always has the form (\ref{etaloc}) up to a common coefficient. In dimensions 
$ d\neq 4 $, the form of (\ref{etaloc}) changes, but remains universal for the fixed $ d $.

Let us now consider the ``global'' viscosity, integrating (\ref{etaloc}) above the stretched horizon
\begin{eqnarray}
\eta_{glob}=\int_{l_c}^{\infty}\eta(\rho)d\rho = \frac{C_T \pi^2}{480 l_c^2}  \,.
\label{etaglob}
\end{eqnarray}
We assume that the local entropy does not depend on $ l_c $. Then, the global entropy, we obtain from (\ref{sres}) and (\ref{cta}) 
\begin{eqnarray}
s_{glob}=\int_{l_c}^{\infty}s(\rho)d\rho = \frac{C_T \pi^3}{120 l_c^2}\,.
\label{sglob}
\end{eqnarray}
The ratio of (\ref{etaglob}) and (\ref{sglob}) saturates the KSS bound
\begin{eqnarray}
\frac{\eta_{glob}}{s_{glob}} = \frac{1}{4\pi}\,.
\label{}
\end{eqnarray}
Thus, we have generalized the result of \cite{Chirco:2010xx, Lapygin:2025zhn} to the case of arbitrary conformal field theory.

Finally, an intriguing conclusion can be drawn by considering the ratio of ``local'' quantities (\ref{etaanom}) and (\ref{sres}). One can see, that this ratio equals $ 3/4\pi $, not $ 1/4\pi $. In the accompanying paper \cite{prep2}, we show that this is not accidental and reflects the general relationship with the speed of sound $ c_s $
\begin{eqnarray}
\frac{\eta}{s} = \frac{1}{4\pi c_s^2}\,.
\label{kssnew}
\end{eqnarray}

%================
\section{Conclusion}
\label{sec_concl}
%================

We explicitly expressed the shear and bulk viscosities of Unruh radiation in terms of the spectral densities $ c^{(2)}(\mu) $ and $ c^{(0)}(\mu) $, and have shown that the unitarity in Minkowski space guarantees thermodynamic irreversibility in Rindler space via the positivity of the viscosities.

In the conformal field theory limit, we have demonstrated the relationship between the shear viscosity and the conformal gravitational anomaly. Thus, ``entanglement viscosity'' is a novel dissipative type of anomalous transport phenomenon, being an effect akin to a ``Cheshire cat's grin'', since a curved space phenomenon manifests itself in flat space. Since phenomenological models predict extreme accelerations in heavy-ion collisions \cite{Prokhorov:2025vak, Karpenko:2018erl, Chernodub:2024wis}, this viscosity could have interesting experimental implications.

We further demonstrate that the integrated viscosity and entropy saturate the KSS bound, thus generalizing earlier predictions. Thereby the Unruh radiation is an example of a fluid with minimal viscosity. 
Moreover, we find that the local viscosity is described by a universal function for any conformal field theory in a fixed number of dimensions.

%=================================================
{\bf Acknowledgements}
%=================================================

The author is thankful to Ioseph L. Buchbinder, Dmitri V. Fursaev, Andrei O. Starinets, Oleg V. Teryaev and Valentin I. Zakharov for stimulating discussions and interest in the work. The work was supported by Russian Science Foundation Grant No. 25-22-00887.

\bibliography{lit}

\begin{thebibliography}{10}

\bibitem{Prigogine1978-uh}
I~Prigogine.
\newblock Time, structure, and fluctuations.
\newblock {\em Science}, 201(4358):777--785, September 1978.

\bibitem{Witten:2024upt}
Edward Witten.
\newblock {Introduction to black hole thermodynamics}.
\newblock {\em Eur. Phys. J. Plus}, 140(5):430, 2025.

\bibitem{Unruh:1976db}
W.~G. Unruh.
\newblock {Notes on black hole evaporation}.
\newblock {\em Phys. Rev.}, D14:870, 1976.

\bibitem{Bisognano:1976za}
J.~J Bisognano and E.~H. Wichmann.
\newblock {On the Duality Condition for Quantum Fields}.
\newblock {\em J. Math. Phys.}, 17:303--321, 1976.

\bibitem{Zamolodchikov:1986gt}
A.~B. Zamolodchikov.
\newblock {Irreversibility of the Flux of the Renormalization Group in a 2D
  Field Theory}.
\newblock {\em JETP Lett.}, 43:730--732, 1986.

\bibitem{Cappelli:1990yc}
Andrea Cappelli, Daniel Friedan, and Jose~I. Latorre.
\newblock {C theorem and spectral representation}.
\newblock {\em Nucl. Phys. B}, 352:616--670, 1991.

\bibitem{Komargodski:2011vj}
Zohar Komargodski and Adam Schwimmer.
\newblock {On Renormalization Group Flows in Four Dimensions}.
\newblock {\em JHEP}, 12:099, 2011.

\bibitem{Damour:1979wya}
Thibaut Damour.
\newblock {\em {Quelques proprietes mecaniques, electromagnet iques,
  thermodynamiques et quantiques des trous noir}}.
\newblock PhD thesis, Paris U., VI-VII, 1979.

\bibitem{Parikh:1997ma}
Maulik Parikh and Frank Wilczek.
\newblock {An Action for black hole membranes}.
\newblock {\em Phys. Rev. D}, 58:064011, 1998.

\bibitem{Kovtun:2004de}
P.~Kovtun, Dan~T. Son, and Andrei~O. Starinets.
\newblock {Viscosity in strongly interacting quantum field theories from black
  hole physics}.
\newblock {\em Phys. Rev. Lett.}, 94:111601, 2005.

\bibitem{Chirco:2010xx}
Goffredo Chirco, Christopher Eling, and Stefano Liberati.
\newblock {The universal viscosity to entropy density ratio from entanglement}.
\newblock {\em Phys. Rev. D}, 82:024010, 2010.

\bibitem{Lapygin:2025zhn}
Dmitry~D. Lapygin, Georgy~Yu. Prokhorov, Oleg~V. Teryaev, and Valentin~I.
  Zakharov.
\newblock {Viscosity, entanglement, and acceleration}.
\newblock {\em Phys. Rev. D}, 112(6):065012, 2025.

\bibitem{landau1987fluid}
L.D. Landau and E.M. Lifshitz.
\newblock {\em Fluid Mechanics: Volume 6}.
\newblock Number v. 6. Butterworth-Heinemann, 1987.

\bibitem{Teryaev:2022mpe}
Oleg Teryaev.
\newblock {Velocity-like maximum polarization: Irreversibility and quantum
  measurements}.
\newblock {\em Phys. Rev. C}, 106(1):L012201, 2022.

\bibitem{Smolkin:2014hba}
Michael Smolkin and Sergey~N. Solodukhin.
\newblock {Correlation functions on conical defects}.
\newblock {\em Phys. Rev. D}, 91(4):044008, 2015.

\bibitem{Solodukhin:2013yha}
Sergey~N. Solodukhin.
\newblock {The a-theorem and entanglement entropy}.
\newblock 4 2013.

\bibitem{Grozdanov:2016fkt}
Sa{\v{s}}o Grozdanov and Andrei~O. Starinets.
\newblock {Second-order transport, quasinormal modes and zero-viscosity limit
  in the Gauss-Bonnet holographic fluid}.
\newblock {\em JHEP}, 03:166, 2017.

\bibitem{Becattini:2017ljh}
F.~Becattini.
\newblock {Thermodynamic equilibrium with acceleration and the Unruh effect}.
\newblock {\em Phys. Rev.}, D97(8):085013, 2018.

\bibitem{Unruh:1983ac}
William~G. Unruh and Nathan Weiss.
\newblock {Acceleration Radiation in Interacting Field Theories}.
\newblock {\em Phys. Rev. D}, 29:1656, 1984.

\bibitem{Osterwalder:1973dx}
Konrad Osterwalder and Robert Schrader.
\newblock {Axioms for euclidean green's functions}.
\newblock {\em Commun. Math. Phys.}, 31:83--112, 1973.

\bibitem{Landau9}
E.~M. Lifshitz and L.~P. Pitaevskii.
\newblock {\em Statistical Physics, Part 2}, volume~9 of {\em Course of
  Theoretical Physics}.
\newblock Butterworth-Heinemann, Oxford, 1980.

\bibitem{Son:2008zz}
Dam~T. Son.
\newblock {Hydrodynamics and gauge/gravity duality}.
\newblock {\em Acta Phys. Polon. B}, 39:3173--3182, 2008.

\bibitem{Laine:2016hma}
Mikko Laine and Aleksi Vuorinen.
\newblock {Basics of Thermal Field Theory}.
\newblock {\em Lect. Notes Phys.}, 925:pp.1--281, 2016.

\bibitem{Kharzeev:2007wb}
Dmitri Kharzeev and Kirill Tuchin.
\newblock {Bulk viscosity of QCD matter near the critical temperature}.
\newblock {\em JHEP}, 09:093, 2008.

\bibitem{Bogoliubov1959-sq}
N~N Bogoliubov and D~V Shirkov.
\newblock {\em Introduction to theory of quantized fields}.
\newblock Monograph \& Texts in Physics \& Astronomical. John Wiley \& Sons,
  Nashville, TN, December 1959.

\bibitem{Solodukhin:2011gn}
Sergey~N. Solodukhin.
\newblock {Entanglement entropy of black holes}.
\newblock {\em Living Rev. Rel.}, 14:8, 2011.

\bibitem{Prudnikov1998-xg}
A~P Prudnikov, Yu~A Brychkov, and O~I Marichev.
\newblock {\em Integrals and series: Special functions vol 2}.
\newblock Taylor \& Francis, London, England, September 1998.

\bibitem{Moschella:2008ik}
Ugo Moschella and Richard Schaeffer.
\newblock {A Note on canonical quantization of fields on a manifold}.
\newblock {\em JCAP}, 02:033, 2009.

\bibitem{Duff:1993wm}
M.~J. Duff.
\newblock {Twenty years of the Weyl anomaly}.
\newblock {\em Class. Quant. Grav.}, 11:1387--1404, 1994.

\bibitem{Brown:1986jy}
M.~R. Brown, A.~C. Ottewill, and Don~N. Page.
\newblock {Conformally Invariant Quantum Field Theory in Static Einstein
  Space-times}.
\newblock {\em Phys. Rev. D}, 33:2840--2850, 1986.

\bibitem{Osborn:1993cr}
H.~Osborn and A.~C. Petkou.
\newblock {Implications of conformal invariance in field theories for general
  dimensions}.
\newblock {\em Annals Phys.}, 231:311--362, 1994.

\bibitem{Kovtun:2008kx}
Pavel Kovtun and Adam Ritz.
\newblock {Universal conductivity and central charges}.
\newblock {\em Phys. Rev. D}, 78:066009, 2008.

\bibitem{Buchel:2008vz}
Alex Buchel, Robert~C. Myers, and Aninda Sinha.
\newblock {Beyond eta/s = 1/4 pi}.
\newblock {\em JHEP}, 03:084, 2009.

\bibitem{Sinha:2009ev}
Aninda Sinha and Robert~C. Myers.
\newblock {The Viscosity bound in string theory}.
\newblock {\em Nucl. Phys. A}, 830:295C--298C, 2009.

\bibitem{Son:2009tf}
Dam~T. Son and Piotr Surowka.
\newblock {Hydrodynamics with Triangle Anomalies}.
\newblock {\em Phys. Rev. Lett.}, 103:191601, 2009.

\bibitem{Fukushima:2008xe}
Kenji Fukushima, Dmitri~E. Kharzeev, and Harmen~J. Warringa.
\newblock {The Chiral Magnetic Effect}.
\newblock {\em Phys. Rev.}, D78:074033, 2008.

\bibitem{Landsteiner:2011cp}
Karl Landsteiner, Eugenio Megias, and Francisco Pena-Benitez.
\newblock {Gravitational Anomaly and Transport}.
\newblock {\em Phys. Rev. Lett.}, 107:021601, 2011.

\bibitem{Prokhorov:2022udo}
G.~Yu. Prokhorov, O.~V. Teryaev, and V.~I. Zakharov.
\newblock {Hydrodynamic Manifestations of Gravitational Chiral Anomaly}.
\newblock {\em Phys. Rev. Lett.}, 129(15):151601, 2022.

\bibitem{Prokhorov:2022snx}
Georgy~Yu. Prokhorov, Oleg~V. Teryaev, and Valentin~I. Zakharov.
\newblock {Gravitational chiral anomaly and anomalous transport for fields with
  spin 3/2}.
\newblock {\em Phys. Lett. B}, 840:137839, 2023.

\bibitem{Prokhorov:2025vak}
G.~Yu. Prokhorov, D.~A. Shohonov, O.~V. Teryaev, N.~S. Tsegelnik, and V.~I.
  Zakharov.
\newblock {Modeling of acceleration in heavy-ion collisions: Occurrence of
  temperature below the Unruh temperature}.
\newblock {\em Phys. Rev. C}, 112(6):064907, 2025.

\bibitem{Diakonov:2025wtt}
Dmitrii Diakonov.
\newblock {De Sitter entropy: On-shell versus off-shell}.
\newblock {\em Phys. Lett. B}, 871:139967, 2025.

\bibitem{Becattini:2023ouz}
Francesco Becattini, Asaad Daher, and Xin-Li Sheng.
\newblock {Entropy current and entropy production in relativistic spin
  hydrodynamics}.
\newblock {\em Phys. Lett. B}, 850:138533, 2024.

\bibitem{Erdmenger:1996yc}
J.~Erdmenger and H.~Osborn.
\newblock {Conserved currents and the energy momentum tensor in conformally
  invariant theories for general dimensions}.
\newblock {\em Nucl. Phys. B}, 483:431--474, 1997.

\bibitem{prep2}
{Causality, the Kovtun-Son-Starinets bound, and a novel sum rule for spectral
  densities}.
\newblock 2026.
\newblock {In preparation}.

\bibitem{Karpenko:2018erl}
Iurii Karpenko and Francesco Becattini.
\newblock {Lambda polarization in heavy ion collisions: from RHIC BES to LHC
  energies}.
\newblock {\em Nucl. Phys.}, A982:519--522, 2019.

\bibitem{Chernodub:2024wis}
M.~N. Chernodub, V.~A. Goy, A.~V. Molochkov, D.~V. Stepanov, and A.~S.
  Pochinok.
\newblock {Extreme Softening of QCD Phase Transition under Weak Acceleration:
  First-Principles Monte~Carlo Results for Gluon Plasma}.
\newblock {\em Phys. Rev. Lett.}, 134(11):111904, 2025.

\end{thebibliography}

\end{document}